\documentclass [prl,superscriptaddress, showpacs, twocolumn]{revtex4-1}
\usepackage{graphicx}
\usepackage{dcolumn}

\makeatletter

\newcommand{\Rmnum}[1]{\expandafter\@slowromancap\romannumeral #1@}
\makeatother

\begin{document}

\title{Field-induced long-range magnetic order in the spin-singlet ground state system YbAl$_3$C$_3$: Neutron diffraction study}

\author{D. D. Khalyavin}
\email{email: dmitry.khalyavin@stfc.ac.uk}
\affiliation{ISIS facility, STFC, Rutherford Appleton Laboratory, Chilton, Didcot, Oxfordshire, OX11-0QX,UK}
\author{D. T. Adroja}
\email{email: devashibhai.adroja@stfc.ac.uk}
\affiliation{ISIS facility, STFC, Rutherford Appleton Laboratory, Chilton, Didcot, Oxfordshire, OX11-0QX,UK}
\author{P. Manuel}
\affiliation{ISIS facility, STFC, Rutherford Appleton Laboratory, Chilton, Didcot, Oxfordshire, OX11-0QX,UK}
\author{A. Daoud-Aladine}
\affiliation{ISIS facility, STFC, Rutherford Appleton Laboratory, Chilton, Didcot, Oxfordshire, OX11-0QX,UK}
\author{M. Kosaka}
\affiliation{Department of Physics, Saitama University, Saitama 338-8570, Japan}
\author{K. Kondo}
\affiliation{Department of Physics, Saitama University, Saitama 338-8570, Japan}
\author{K. A. McEwen}
\affiliation{London Centre for Nanotechnology, and Department of Physics and Astronomy, University College London, Gower Street, London WC1E 6BT, UK}
\author{J. H. Pixley}
\affiliation{Department of Physics and Astronomy, Rice University, Houston, Texas, 77005, USA}
\date{\today}
\author{Qimiao Si}
\affiliation{Department of Physics and Astronomy, Rice University, Houston, Texas, 77005, USA}
\date{\today}

\begin{abstract}
The $4f$-electron system YbAl$_3$C$_3$ with a non-magnetic spin-dimer ground state has been studied by neutron diffraction in an applied magnetic field. A long-range magnetic order involving both ferromagnetic and antiferromagnetic components has been revealed above the critical field H$_C\sim $ 6T at temperature T=0.05K. The magnetic structure indicates that the geometrical frustration of the prototype hexagonal lattice is not fully relieved in the low-temperature orthorhombic phase. The suppression of magnetic ordering by the remanent frustration is the key factor stabilizing the non-magnetic singlet ground state in zero field. Temperature dependent measurements in the applied field H=12T  revealed that the long-range ordering persists up to temperatures significantly higher than the spin gap indicating that this phase is not directly related to the singlet-triplet excitation. Combining our neutron diffraction results with the previously published phase diagram, we support the existence of an intermediate disordered phase as the first excitation from the non-magnetic singlet ground state. Based on our results, we propose YbAl$_3$C$_3$ as a new material for studying the quantum phase transitions of heavy-fermion metals under the influence of geometrical frustration.
\end{abstract}

\pacs{75.25.-j}

\maketitle

\indent The formation of a non-magnetic singlet ground state due to spin dimerization is an extremely rare phenomenon in $4f$-electron systems. It requires strong quantum effects which are usually significant only in low dimensional spin $S=1/2$ systems. Due to a larger total angular momentum $J$, and the three-dimensional character of interactions via conduction electrons, the spin dimer state is not favourable in most rare earth compounds. The exceptional cases are Yb$_4$As$_3$ where an effective spin $S=1/2$ Heisenberg chain is believed to be realized \cite{ref:1} and the recently proposed spin-dimer system YbAl$_3$C$_3$ \cite{ref:2,ref:3,ref:3a,ref:4}. Due to the effect of the crystalline electric field, the ground state Kramers doublet is suggested to be well isolated from the exited states in these compounds and the $4f$ electrons at low temperature are expected to behave as a $S=1/2$ spin system.\\
\indent The non-magnetic nature of the ground state in YbAl$_3$C$_3$ was initially interpreted as an antiferroquadrupolar ordered state \cite{ref:5} which takes place at T$_S\sim $80K, where the specific heat exhibits a $\lambda $-type anomaly. Later, this interpretation was discarded by Ochiai et al. \cite{ref:2} who observed a similar sharp peak in the specific heat for the Lu-based analogue at 110 K and attributed it to a structural phase transition present in both compounds. The new interpretation proposed by Ochiai et al. \cite{ref:2} implies formation of isolated dimers due to the structural distortions which promote the non-magnetic spin-singlet ground state. This idea explains the low temperature specific heat and magnetization data assuming the spin gap to be $\sim $15K. More direct evidence of the singlet-triplet excitations in YbAl$_3$C$_3$ has been presented by Kato et al. \cite{ref:3} and Adroja et al. \cite{ref:3a} based on inelastic neutron scattering.\\
\indent By analogy with insulating $d$-electron dimer systems, one can expect field-induced long-range magnetic order in YbAl$_3$C$_3$ at the critical magnetic field closing the spin-gap due to the Zeeman effect. It has been shown that this kind of quantum phase transition can be modeled as a Bose-Einstein condensation in a system of weakly interacting bosons \cite{ref:6,ref:7}. At the same time, the metallic nature of YbAl$_3$C$_3$  introduces an interplay between RKKY and Kondo interactions which, in turn, suggests that the magnetic field will induce behavior that is distinct from that in insulating dimer systems. The aim of the present study, therefore, was to directly seek and investigate the field-induced long-range magnetic order in YbAl$_3$C$_3$ through neutron diffraction. Our data indeed reveal a change of the ground state from non-magnetic below the critical field  H$_C$=6T to magnetically ordered above H$_C$. Surprisingly, the suggested magnetic structure does not imply a strong exchange coupling for the expected isolated dimers and indicates the presence of geometrical frustration in spite of the lack of the threefold symmetry in the low temperature structural phase. Based on the obtained results, one can conclude that the remanent frustration is the crucial ingredient promoting the low-temperature singlet ground state and is responsible for the unusual excitations in the system; in turn, our results suggest that YbAl$_3$C$_3$ provides a new setting to study heavy-fermion quantum phase transitions under the influence of geometrical frustration.\\
\begin{figure}[t]
\includegraphics[scale=1.0]{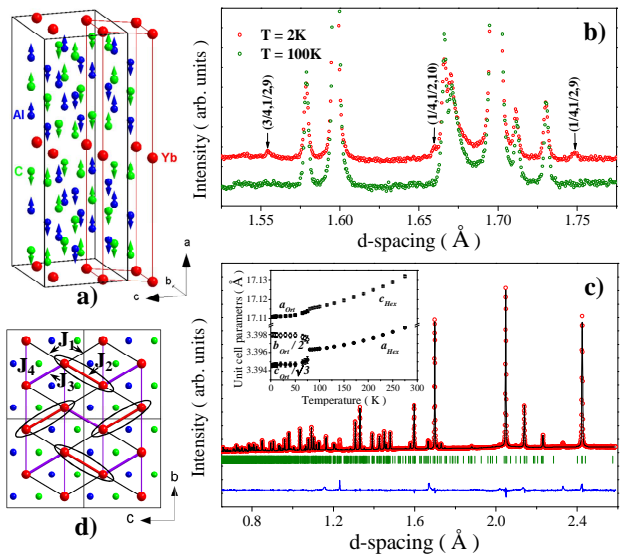}
\caption{(Color online) (a) Orthorhombic structure of YbAl$_3$C$_3$ (the parent hexagonal cell is shown as well). The primary orthorhombic displacements of Al and C are represented by arrows. (b) Neutron diffraction patterns collected above and below the structural phase transition. (c) Rietveld refinement of the neutron diffraction pattern collected at T=2K. Inset shows the unit cell parameters as a function of temperature. (d) Triangular layer formed by Yb$^{3+}$ ions and the exchange parameters in the orthorhombic phase. The interatomic Yb-Yb distances (evaluated at T=2K) corresponding to the non-equivalent exchange parameters are: 3.3970(1)$\AA$ for J$_i(i=1-3)$ and 3.3995(1) for J$_4$. J$_2$-bonds are arbitrarily taken to show the isolated dimers as ellipses around the bonds.}
\label{fig:F1}
\end{figure}
\indent The high resolution neutron diffraction experiments were performed on a 2g polycrystalline sample YbAl$_3$C$_3$ prepared as described by Kosaka et al. \cite{ref:5} The measurements were carried out on HRPD (in zero magnetic field and 2K$\leq$T$\leq$300K temperature range) and the WISH diffractometer \cite{ref:7a} (in the 0T$\leq$H$\leq$12T field range at T=0.05K and in the 0.05T$\leq$T$\leq$20K temperature range at H=12T) at the ISIS Facility of the Rutherford Appleton Laboratory (UK). To avoid alignment of sample grains, in the field-dependent measurements, the powder was cold pressed into pellets fitting tightly inside a 6mm vanadium can. The refinement of the crystal and magnetic structures were done using the FullProf software \cite{ref:8}.\\ 
\indent The high temperature neutron diffraction patterns (T$>$80K) can be successfully refined in the hexagonal $P6_3/mmc$ space group consistent with the previous diffraction studies \cite{ref:5,ref:9}. The structural parameters obtained at T=100K are summarised in Table {\ref{table:T1}. The Yb$^{3+}$ ions create two-dimensional triangular layers (Fig. \ref{fig:F1}a) causing the geometrical frustration for the associated $4f$ magnetic moments. The low dimensionality of the system is ensured by the large inter-layer distance $\sim 8.6\AA$ resulting in a significant predominance of the in-plane interactions over the out-of-plane ones. Below T$_S$, a set of very weak additional reflections appears (Fig. \ref{fig:F1}b), indicating the structural phase transition. In agreement with the single crystal X-ray diffraction work by Matsumura et al. \cite{ref:10}, these reflections can be indexed with one of the symmetry related propagation vectors: ${\bf k_1}=1/4,1/4,0$, ${\bf k_2}=-1/4,1/2,0$ or ${\bf k_3}=-1/2,1/4,0$. However, the atomic coordinates reported by these authors did not work with our diffraction data. 
To get a better structural model a detailed symmetry analysis has been performed. The above mentioned wave vector star combines four irreducible representations, $\Lambda _i (i=1-4)$. For each representation, isotropy subgroups and displacive modes were generated using ISOTROPY \cite{ref:11} and ISODISTORT \cite{ref:12} software and checked in the refinement procedure versus the experimentally measured superstructure reflections. We restricted our analysis by considering only subgroups associated with a single ${\bf k}$ and ${\bf -k}$ pair as experimentally found by Matsumura et al. \cite{ref:10}. 
\begin{table}[t]
\caption{Structural parameters of YbAl$_3$C$_3$ refined in the hexagonal ($P6_3/mmc$, $a$=3.39804(1)$\AA$, $c$=17.12430(9)$\AA$, $R_{Bragg}$=4.18 $\%$) and orthorhombic ($Pbca$, $a$=17.11895(9)$\AA$, $b$=6.79900(10)$\AA$, $c$=5.88243(8)$\AA$, $R_{Bragg}$=4.49 $\%$) phases.}
\centering 
\begin{tabular*}{0.48\textwidth}{@{\extracolsep{\fill}} c c c c c c} 
\hline\hline\\ 
Atom & Site & $x$ & $y$ & $z$ & $B_{iso}$ \\ [1.5ex] 
\hline\\ 
 &  & $P6_3/mmc$ &  T=100K &   & \\ [1.5ex] 
\hline\\  
Yb & 2$a$ & 0 & 0 & 0 & 0.92(2)\\ 
Al1 & 4$f$ & 1/3 & 2/3 & 0.13228(9) & 0.87(3)\\
Al2 & 2$d$ & 1/3 & 2/3 & 3/4 & 1.65(5)\\
C1 & 4$f$ & 1/3 & 2/3 & 0.59057(5) & 1.19(2)\\
C2 & 2$c$ & 1/3 & 2/3 & 1/4 & 1.19(2)\\[1.5ex]
\hline\\ 
 &  & $Pbca$ &  T=2K &   & \\[1.5ex] 
\hline\\ 
Yb & 8$a$ & 0 & 7/8 & 3/4 & 0.85(2)\\ 
Al11 & 8$a$ & 0.13297(43) & 1/8 & 11/12 & 0.92(3)\\
Al12 & 8$a$ & 0.13159(43) & 5/8 & 11/12 & 0.92(3)\\
Al2 & 8$a$ & 0.76017(20) & 1/8 & 11/12 & 1.20(5)\\
C11 & 8$a$ & 0.58585(13) & 1/8 & 11/12 & 1.03(2)\\
C12 & 8$a$ & 0.59529(13) & 5/8 & 11/12 & 1.03(2)\\
C2 & 8$a$ & 0.25023(28) & 1/8 & 11/12 & 0.96(2)\\
\\
\hline
\hline  
\end{tabular*}
\label{table:T1} 
\end{table}
The analysis resulted in the conclusion that the displacive modes involving only Al and C (mainly Al2 and C1) and associated with the $\Lambda _3$ representation and the $(a,0,0,a,0,0)$ order parameter direction drive the phase transition at T$_S$. These primary modes have the $Pbca$ symmetry in agreement with the reflection conditions deduced by Matsumura et al. \cite{ref:10} and induce the atomic displacements along the hexagonal $c$-axis (Fig. \ref{fig:F1}a). We found the Yb ions were not displaced within the precision of our diffraction experiment. Taking into account this result the final refinement (Fig. \ref{fig:F1}c) was done in the orthorhombic $Pbca$ space group (related to the hexagonal $P6_3/mmc$ by: ${\bf a_o}={\bf c_h}$, ${\bf b_o}=2{\bf a_h}+2{\bf b_h}$ and ${\bf c_o}={\bf b_h}-{\bf a_h}$) using only four parameters varying the $x$ coordinates for the six nominally independent Al and C sites. The $y$ and $z$ coordinates for these atoms as well as all coordinates for Yb were fixed to their high symmetry values corresponding to the high-temperature hexagonal phase (Table \ref{table:T1}). This simplified approach to the refinement of the low-temperature phase is the only one possible due to the limited number of superstructure reflections in the powder data allowing us to determine only primary distortions.\\
\begin{figure}[t]
\includegraphics[scale=1.05]{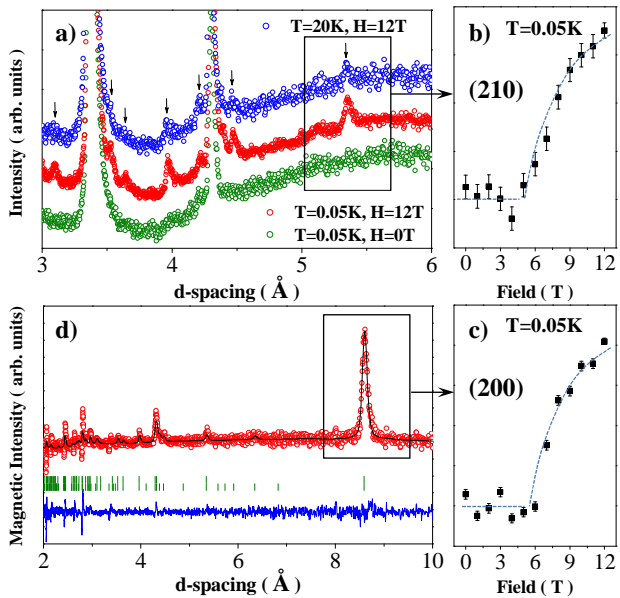}
\caption{(Color online) (a) Neutron diffraction patterns collected at different temperatures and magnetic fields. Arrows indicate positions of the additional reflections induced by the magnetic field. (b) Integrated intensity of the antiferromagnetic peak (210) as a function of the magnetic field. (c) Ferromagnetic contribution to the (200) nuclear reflection as a function of the magnetic field. (d) Refinement of the magnetic intensities obtained by subtraction of the patterns collected at T=0.05K in H=12T and H=0T (magnetic Bragg factor $R_{Bragg}$=6.2$\%$).}
\label{fig:F2}
\end{figure}
\indent Considering the effect of the symmetry lowering on the exchange topology of the Yb sublattice, the favourable symmetry conditions to form isolated dimers should be pointed out. In the low temperature orthorhombic phase there are four non-equivalent in-plane exchange couplings shown in Figure \ref{fig:F1}d. $J_2$ and $J_3$ form isolated dimers and can potentially promote the non-magnetic singlet ground state. The corresponding Yb-Yb inter-atomic distances vary only slightly due to the small variations of the $a$ and $b$ unit cell parameters (Fig. \ref{fig:F1}c inset), which indicates that the main factor renormalizing the exchange parameters should be related to the Al and C displacements.\\
\begin{figure}[t]
\includegraphics[scale=1.00]{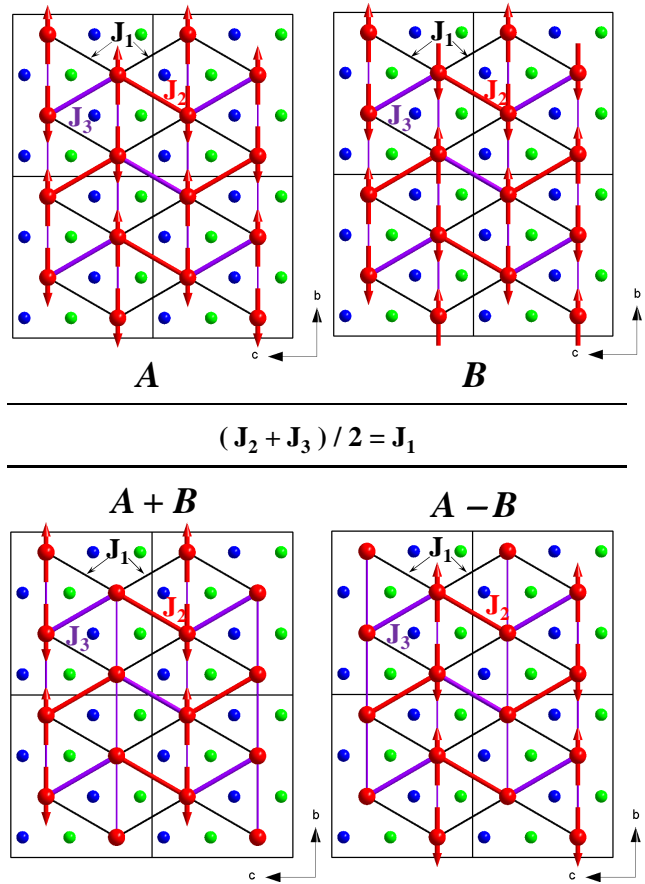}
\caption{(Color online) (Top) Two degenerate spin configurations denoted as $A$ and $B$ in a triangular layer. (Bottom) Superposition of the degenerate configurations taking with same (left) and opposite signs (right).}
\label{fig:F3}
\end{figure}
\indent The absence of any sign of long-range magnetic ordering in our diffraction data down to 0.05K is fully consistent with the previous neutron diffraction \cite{ref:5}, M{\"o}ssbauer spectroscopy \cite{ref:13} and muon spin relaxation \cite{ref:14} studies. However, application of magnetic field $\sim$ 6T changes the non-magnetic ground state of YbAl$_3$C$_3$. A clear ferromagnetic signal on top of some nuclear peaks and a set of additional reflections in a low-$Q$ region (Fig. \ref{fig:F2}a) indicate the onset of long-range magnetic order. The field dependence of both ferromagnetic (Fig. \ref{fig:F2}c) and antiferromagnetic (Fig. \ref{fig:F2}b) components demonstrates a critical behaviour with the critical exponents 0.24(3) and 0.32(6), respectively. The quantitative refinement of the magnetic intensity (Fig. \ref{fig:F2}d) revealed a uniform ferromagnetic component along the orthorhombic $c$-axis and an antiferromagnetic component along the $b$-axis. The latter does not require enlargement of the orthorhombic nuclear unit cell and all magnetic peaks can be indexed with the ${\bf k}=0$ propagation vector.
Using representation theory \cite{ref:11,ref:12}, we classified different magnetic configurations according to irreducible representations of the $Pbca$ space group and used them in the refinement procedure. No solution within a single irreducible representation has been found, therefore combinations of different representations were tested. We found that many combinations describe the observed magnetic intensities equally well (Fig. \ref{fig:F2}d) and no unique solution can be deduced from the powder data. However, the important point is that all these solutions can be presented as admixtures of two configurations denoted as $A$ and $B$ in Figure \ref{fig:F3} (top) and representing magnetic ordering in the triangular layers. It can be a simple interchanging of these layers, $[A]_{x\sim 0} \rightarrow  [B]_{x\sim 0.5} \rightarrow  [A]_{x\sim 1} \rightarrow \cdots$ along the orthorhombic $a$-axis (the former hexagonal $c$-axis), which in combination with the ferromagnetic component results in the canted structure (Fig. \ref{fig:F4}a) with monoclinic $P2_1'$ symmetry. Or a more complex orthorhombic $P2_1'2_1'2_1$ combination involving disordered antiferromagnetic components on half of the Yb sites (Fig. \ref{fig:F3} bottom) and implying $[A-B]_{x\sim 0} \rightarrow  [A+B]_{x\sim 0.5} \rightarrow  [A-B]_{x\sim 1} \rightarrow  \cdots$ alternation (Fig. \ref{fig:F4}b).
\begin{figure}[t]
\includegraphics[scale=1.0]{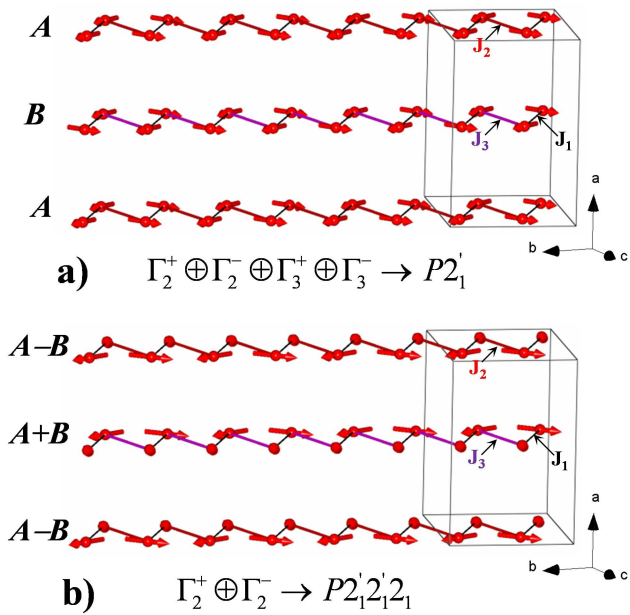}
\caption{(Color online) Two examples of the field-induced magnetic structures combining degenerate $A$ and $B$ triangular layers. The monoclinic $P2_1'$ structure (a) is fully ordered, the ferromagnetic (along the $c$-axis) and antiferromagnetic (along the $b$-axis) components equal 1.6(2)$\mu_B$ and 1.6(2)$\mu_B$, respectively, in the magnetic field H=12T. The orthorhombic $P2_1'2_1'2_1$ structure (b) consists of fully ordered ferromagnetic component, 1.6(2)$\mu_B$, and partially disordered antiferromagnetic one, 2.2(2)$\mu_B$ (only half of the sites carry non-zero antiferromagnetic component).}
\label{fig:F4}
\end{figure}
Practically, it means interchanging two types of the triangular layers; one of them is randomly represented by the $A$ or $B$ configurations taken with different signs, the second one is randomly represented by these configurations taken with same sign. Note that the $[B-A]_{x\sim 0} \rightarrow  [A+B]_{x\sim 0.5} \rightarrow  [B-A]_{x\sim 1} \rightarrow  \cdots$ stacking type is not equivalent to the previous one and has a distinct magnetic symmetry but reproduces equally well the diffraction pattern. 
There are many intermediate situations with sites half fully-ordered and half partially-ordered (intermediate between the structures shown in Figure \ref{fig:F4}a and \ref{fig:F4}b). However, the most symmetric variant (Fig. \ref{fig:F4}b) is considered to be preferable since it involves only two representations and therefore a lower degree of degeneracy. The value of the ordered moments for Yb$^{3+}$ ions depends on the model and varies from 2.3(3)$\mu_B$ for the fully-ordered configurations up to 2.7(3)$\mu_B$ for the partially-disordered ones with only half of the sites carrying the ordered antiferromagnetic component.\\
\indent The admixture of the $A$ and $B$ configurations signifies their degeneracy and points to the fact that the geometrical frustration is not fully relieved by the structural phase transition at T$_S$. Quantitatively, the frustration can be expressed by the ratio $(J_2+J_3)/2=J_1$ between the in-plane exchange parameters, which is held in the orthorhombic phase. This ratio is not consistent with the strong predominance of either $J_2$ or $J_3$ exchange coupling expected from the symmetry arguments to form isolated dimers. A possible explanation of this experimental fact is that the ground state of Yb$^{3+}$ ions in the field-induced ordered phase is essentially different from the zero-field singlet state. This assumption comes from the large ordered moment of Yb obtained from the experimentally measured magnetic intensities. The new ground state adopts the bigger moment to gain full advantage of the magnetically ordered state and can renormalize the exchange parameters in the system.\\
\indent The results obtained indicate directly that the frustrated nature of the orthorhombic phase is the key factor to stabilize the non-magnetic singlet ground state in YbAl$_3$C$_3$ as it has been originally suggested by Kato et al. \cite{ref:3} The magnetic fluctuations between degenerate manifolds caused by the frustration prevents the system from choosing a unique ordered pattern and the spin-dimerization takes place to lift the degeneracy of the ground state.\\
\indent Field-induced transitions to magnetically ordered phases are commonly observed in many $d$-electron dimer systems \cite{ref:7}. However, in strong contrast with these systems, YbAl$_3$C$_3$ does not show a cusp in magnetization curves measured in fields H$>$H$_C$ as a function of temperature \cite{ref:4}, which can be attributed to the onset of magnetic ordering at the critical temperature T$_N$(H). To clarify this important issue, we performed temperature dependent diffraction measurements in the magnetic field H=12T applied at 0.05K. Surprisingly, both the ferromagnetic and antiferromagnetic contributions to the Bragg peaks were observed up to the highest measured temperature T=20K (Fig. \ref{fig:F2}a). A detailed temperature study in different magnetic fields is in progress and will be reported elsewhere. This observation indicates that the field-induced long-range magnetic order is not directly related to the singlet-triplet excitation and can be induced at temperatures significantly higher than the spin gap in agreement with the magnetization data recently reported by Hara et al. \cite{ref:4}\\
\indent Taking into account the frustrated nature of the YbAl$_3$C$_3$ crystal structure and the strong antiferromagnetic interactions (the paramagnetic Curie temperature is $\sim $-100K), the system is expected to be in a "classical" spin-liquid regime in a wide temperature range below T$_S$. Thus, the field-induced long-range magnetic order can be associated with the properties of this phase rather than with the singlet-triplet excitation. This consideration presumes that the spin dimerization takes place from the disordered spin-liquid phase, in other words, the non-magnetic quantum singlet state competes with this phase and therefore, one can expect that the first excitation, when the spin gap is closed, is the disordered spin-liquid phase as well. The long-range ordering therefore is the second excited state. The presence of the intermediate field-induced disordered phase at low temperatures cannot be directly deduced from the powder diffraction data since this phase does not contribute to the Bragg diffraction, but the recent single crystal magnetization data reported by Hara et al. \cite{ref:4} clearly indicates the presence of the intermediate phase below 6T.  This intermediate phase indeed has been recently confirmed through inelastic neutron scattering study in applied magnetic field by Adroja et al. \cite{ref:3a}, which revealed that the position and intensity of the inelastic excitations changed dramatically between H=3 and 5T.\\
\indent The metallic nature of YbAl$_3$C$_3$ suggests the importance of Kondo effect and its interplay with RKKY interactions. We therefore expect it to be a new material that can be used to study the global phase diagram of antiferromagnetic heavy-fermion metals \cite{ref:15,ref:16}. There is a burst of recent interest in studying such a phase diagram through materials that can tune the degree of local-moment quantum fluctuations through dimensionality \cite{ref:17} or geometrical frustration \cite{ref:18, ref:19,ref:20}. YbAl$_3$C$_3$ belongs to the latter class of materials, with the distinction that the relevance of geometrical frustration is explicitly established even though the system is metallic. Further support for our picture comes from the observation of a large value of the Sommerfeld coefficient ($\gamma = C/T$) \cite{ref:4} at intermediate magnetic fields.\\
\indent In conclusion, the geometrically frustrated system YbAl$_3$C$_3$ exhibits a first order structural phase transition from hexagonal $P6_3/mmc$ to orthorhombic $Pbca$ symmetry at T$_S \sim$80K. The primary order parameter driving the transition involves displacements of Al and C ions along the hexagonal $c$-axis. The structural distortions renormalize the exchange parameters promoting the formation of isolated dimers but the orthorhombic symmetry does not fully release the frustration. Suppression of magnetic ordering by the remanent frustration stabilizes the non-magnetic singlet ground state. Application of a magnetic field above H$_C\sim$6T induces long-range magnetic order at T=0.05K. The magnetic structure involves a homogeneous ferromagnetic component along the orthorhombic $c$-axis and an antiferromagnetic component along the $b$-axis. The latter is likely to be disordered on half of the Yb sites. In the magnetic field H=12T, both the ferromagnetic and antiferromagnetic components persist up to 20K indicating that the long-range order is not directly related to the singlet-triplet excitation. A field-induced intermediate disordered phase is likely to exist as the first excitation from the non-magnetic singlet ground state. This opens the primary question whether the long-range magnetically ordered phase in YbAl$_3$C$_3$ really displays similar physics to that of a Bose-Einstein condensation of magnons and how the Kondo effect interplays with the magnetic frustration. \\
\indent Acknowledgment: We would like to thank A.D. Hillier, J.-G. Park, J.R. Stewart, T. Guidi and P. Riseborough for interesting discussions. DTA would like to acknowledge financial assistance from CMPC-STFC Grant No. CMPC-09108. The work at Rice has been supported in part by the NSF Grant No. DMR-1006985 and the Robert A. Welch Foundation Grant No. C-1411.

\end{document}